\documentclass{PoS}

\usepackage{amsmath}
\usepackage{graphicx}
\usepackage{amssymb}
\usepackage{subfigure}
\usepackage{cancel}
\usepackage{relsize}
\usepackage{bbold}
\usepackage{lineno}
\usepackage{braket}
\usepackage{slashed}
\usepackage{multirow}
\usepackage{placeins}
\usepackage{xspace}

\def\babar{\mbox{\slshape B\kern-0.1em{\smaller A}\kern-0.1em
    B\kern-0.1em{\smaller A\kern-0.2em R}}}

\def\Kbar{{\kern 0.2em\overline{\kern -0.2em \ensuremath{K}}{}}\xspace}
\newcommand\Kzb{{\ensuremath{\Kbar{}^0}}\xspace}
\newcommand{\CP}{{\ensuremath{C\!P}}\xspace}
\newcommand{\Dz}{\ensuremath{D^0}\xspace}
\def\Dbar{{\kern 0.2em\overline{\kern -0.2em\ensuremath{D}}{}}\xspace}
\newcommand{\Dzb}{\ensuremath{\Dbar{}^0}\xspace}
\def\Bbar{{\ensuremath{\kern 0.18em\overline{\kern -0.18em \ensuremath{B}}{}}}\xspace}
\newcommand{\stat}{\ensuremath{(\mathrm{stat.})}}
\newcommand{\syst}{\ensuremath{(\mathrm{syst.})}}
\newcommand{\eg}{\mbox{\itshape e.g.}\xspace}
\newcommand{\DzDzb}{\Dz--\Dzb}

\title{Phenomenological and Experimental Developments in Charm Physics: The WG7 Report from CKM 2016}

\ShortTitle{Phenomenological and Experimental Developments in Charm Physics}

\author{Giulia Casarosa\\
        Institut f\"ur Kernphysik, Johannes Gutenberg-Universit\"at Mainz,\\
        Johann-Joachim-Becher-Weg 45 D 55128 Mainz, Germany\\
        E-mail: \email{casarosa@uni-mainz.de}}

\author{Angelo Di Canto\\
        European Organization for Nuclear Research (CERN),\\
        CH-1211 Geneva 23, Switzerland\\
        E-mail: \email{angelo.di.canto@cern.ch}}

\author{\speaker{Ayan Paul}\\
        INFN, Sezione di Roma, Piazzale A. Moro 2, I-00185 Roma, Italy.\\
        E-mail: \email{ayan.paul@roma1.infn.it}}

\abstract{We present an overview of recent developments in charm physics reported in the Working~Group~7 meetings of the CKM 2016 workshop. Progresses in experimental analysis and techniques were reported by LHCb, Belle, \babar\ and BESIII along with projections for the future. Developments were also reported in the phenomenological extraction of both direct and indirect \CP violation from data in two-, three- and four-body hadronic decays of the $D$ mesonic system.}

\FullConference{9th International Workshop on the CKM Unitarity Triangle\\
		28  November - 3 December 2016\\
		Tata Institute for Fundamental Research (TIFR), Mumbai, India}

\begin{document}

%%%%%%%%%%%%%%
\section{Introduction}
\label{sec:intro}
%%%%%%%%%%%%%%

While the LHC ramps up data through Run\,2, and Belle\,II prepares for the commencement of its run in the near future, it is important to examine the prospects that the study of charm dynamics brings about in our understanding of fundamental physics at the GeV scale. Physics at the charm threshold has the dual aspect of being in the psuedo-perturbative regime and being a platform for probing dynamics beyond the Standard Model (SM) since the latter leaves very tiny traces in quite a few decay modes. While charm dynamics has been a second cousin in the study of flavour physics because of its detachment from both heavy quark techniques and chiral perturbation theory and not been given proportionate attention in the realms of experimental studies, much development has been made in the past few years with renewed interests in phenomenological and experimental studies. There are significant hurdles in the development of theoretical frameworks for charm dynamics which lie beyond perturbation theory.
\begin{itemize}
\itemsep0em 
\item Traditional techniques for heavy quarks: factorization, pQCD etc. do not work, even conceptually.
\item Attempts have been made to use hybrid forms of chiral perturbation theory but charm is too massive for that.
\item Attempts have been made to use $SU(3)_{FL}$ breaking arguments in the $(u, d, s)$ multiplets to extract reduced matrix elements, analyze tree and penguin contributions etc. However, traditional $SU(3)_{FL}$ breaking through the strange quark mass insertion lies on shaky ground as some assumptions have to be made.
\item When one uses the OPE approach, there is no clear distinction between the long distance and the short distance due to the dominance of light quark operators.
\item Charm hadrons are heavy enough to decay into high multiplicity states, a bane and a boon.
\end{itemize}

Experiments have been been designing rich programs for charm dynamics which should yield interesting results in the near future:
\begin{itemize}
\itemsep0em 
\item Charm is being produced in significantly large numbers, especially in hadron machines: at LHCb  $\sim O(10^{13})$ charm pairs are being produced and the future holds promises of much more.
\item Extraction of charm signals from the background has significantly improved.
\item A beauty factory can produce as much charm as beauty.
\item Dedicated charm threshold runs possible at super flavour factories.
\item BESIII provides important input that can be used to analyze charm data at the LHCb~\cite{Malde:2223391} and Belle\,II.
\end{itemize}

In the following section we will review some key phenomenological developments in charm dynamic followed by details of the results from LHCb, Belle, \babar\ and BESIII along with future experimental prospects.

%%%%%%%%%%%%%%
\section{Phenomenological Developments}
\label{sec:pheno}
%%%%%%%%%%%%%%

Both the study of strong dynamics and the probe for new physics in charm dynamics lays great focus on direct and indirect \CP violation,  in the decay and oscillation of charmed mesons respectively. In this review, we will focus on phenomenological developments in the estimations of direct \CP asymmetries in the hadronic decays, on \DzDzb oscillations and possibilities of measuring indirect \CP violation in multibody hadronic decays through amplitude analyses.

%%%%%%%%%%%%%%
\subsection{\CP violation in two-body hadronic decays of the \boldmath $D$ system}
\label{sec:pheno:twobody}
%%%%%%%%%%%%%%
In the past few years there has been some renewed interest in developing parametric models based on topological or $SU(3)_{FL}$ arguments to use experimental measurements of the branching fraction of $D\to P P $ decays ($D = D^0, D^\pm_{(s)}$ and $P=\pi,K,\eta,\eta^\prime$) to get a handle on \CP asymmetries in these channels. The obvious hurdle in the estimation of the latter has been the the lack of theoretical control on the crucial Penguin amplitude, the magnitude and phase of which quantifies the \CP asymmetry in these channels. In addition, the complexities of disentangling the short distance from the long distance at scales of these decays render theoretical estimation of many other topologies difficult. While it is well known that direct \CP asymmetries can primarily find expression in the singly Cabibbo suppressed (SCS) modes the measured branching fractions of the Cabbibo favoured (CF) and doubly Cabbibo suppressed (DCS) modes can be used to constrain the parameters in the \CP conserving part of the amplitude~\cite{Hiller:2012xm,Muller:2015rna,Nierste:2015zra}. 

U. Nierste presented a study on the estimation of \CP violation in two-body decays of the charmed meson system focussing on $D\to P P$ decays. The first part~\cite{Muller:2015rna} addresses how sum rules for \CP asymmetries can be built amongst a subset of these channels to eliminate the necessity for estimating the Penguin amplitudes which are difficult to predict within the SM. The other topologies appearing in the \CP conserving part are extracted from branching fraction measurements using linear $SU(3)_{FL}$ breaking arguments and $1/N_c$ counting as explained in~\cite{Muller:2015lua}. The sum rules connect two sets of three asymmetries $(D^0\to K^+K^-,\; D^0\to \pi^+\pi^-,\;D^0\to\pi^0\pi^0)$ and $(D^+\to \Kzb K^+,\;D^+_s\to K^0\pi^+,\;D^+_s\to K^+\pi^0)$. This improves the predictability of these asymmetries significantly but depends on the assumed size of $SU(3)_{FL}$ breaking.

Recently, both LHCb~\cite{Aaij:2015fua} and Belle~\cite{Abdesselam:2016gqq} presented their results on $A_{\CP}(D^0\to K^0_SK^0_S)$:
\begin{eqnarray}
A^{\rm LHCb}_{\CP}(D^0\to K^0_SK^0_S)&=&(-2.9 \pm 5.2\stat\pm 2.2\syst)\% \nonumber\\
A^{\rm Belle}_{\CP}(D^0\to K^0_SK^0_S)&=&(0.02 \pm 1.53\stat\pm 0.17\syst)\%\nonumber
\end{eqnarray}
In the SM at the $SU(3)_{FL}$ conserving limit the branching fraction of this mode vanishes. Hence the \CP asymmetry is enhanced due to the suppression of the \CP conserving amplitude by this broken symmetry. The \CP violating part of the amplitude receives, possibly sizable contribution from the tree-level exchange diagram. The authors of~\cite{Nierste:2015zra} estimate the SM contribution to be:
\begin{equation}
A_{\CP}(D^0\to K^0_SK^0_S)\le 1.1\%\;\;@\;\;95\%\;\;{\rm CL}\nonumber
\end{equation}
An overview of these results can be found in~\cite{Nierste:2017hsw}.
%%%%%%%%%%%%%%
\subsection{Multi-body hadronic decays of the \boldmath $D$ system}
\label{sec:pheno:multi}
%%%%%%%%%%%%%%
Amplitude analysis is a good way to study interference effects and hence critical for the study of strong phases and \CP violation. However, the extraction of information from Dalitz plots are often model dependent, a review of which can be found in~\cite{Amato:2016xjv}. J. Rademacker presented model comparisons in Dalitz plot analysis done for $D^+\to K^-K^+K^+$ in~\cite{LHCb-CONF-2016-008} where three different isobar models were used with different components in the high $K^+K^-$ mass region including a model with a non-resonant component which proved disfavoured compared to the other two. These models were instrumental in this study of the $K^+K^-$ S-wave properties which is the dominant component in this decay mode like other three body decay modes with a pair of identical particles in the final state.

Another analysis done for $D^0\to K^0_SK^-\pi^+$ and $D^0\to K^0_SK^+\pi^-$ was presented based on~\cite{Aaij:2015lsa} where GLASS and LASS parameterizations of the $K\pi$ S-wave were used. Both models proved to give good representation of the data indicating absence of strong dependence of the analysis on the S-wave parametrization. A \CP violation search was also performed and no significant effects were found. These models will be useful in future studies of $D^0 -\bar{D}^0$ mixing, indirect \CP violations and the CKM angle $\gamma$. Several studies of \CP violation in tree body decay modes were also presented from Belle~\cite{Peng:2014oda} and CDF~\cite{Aaltonen:2012nd} for $D^0\to K_S\pi^+\pi^-$ and \babar~\cite{Lees:2012nn} for $D^\pm \to K^+K^-\pi^\pm$. Emphasis was laid on input from the charm threshold data from CLEO-c and BESIII which can lead to model independent analyses such as the one done for $D^0\to K_S\pi^+\pi^-$ at LHCb~\cite{Aaij:2015xoa}

For four body decays such as $D^0\to\pi^+\pi^-\pi^+\pi^-$, a five dimensional generalization of the Dalitz plot can be made as studied in~\cite{dArgent:2016rbp} using data from CLEO-c. Due to the complexity of the intermediate resonances, a large number of components have to be used, 18 in this case. No \CP violation is found in this study. It was also emphasized that four body decay modes can provide new observables unavailable in 3 body modes such as T-odd moments which can then be used to construct measures of \CP violation from the difference of T-odd moments in \CP conjugate states. This was studied by LHCb~\cite{Aaij:2014qwa} and \babar~\cite{delAmoSanchez:2010xj} for $D^0 \to K^+K^-\pi^+\pi^-$ decays. Neither of the studies found evidence of \CP violation.  

%%%%%%%%%%%%%%
\subsection{UTfit results in oscillations and \CP violation in the neutral \boldmath  $D$ system}
\label{sec:pheno:utfit}
%%%%%%%%%%%%%%

Oscillations of the neutral $D$ system are well established. However, no signatures of \CP violation from oscillations have been traced as yet. The presence of indirect \CP violation would be implied by
\begin{equation}
\left|\frac{q}{p}\right| \ne 1\;\;\text{and/or}\;\;\phi=\arg\left(\frac{q}{p}\right)\ne0,\pi.
\end{equation}
The UTfit Collaboration performs a fit for the mixing parameters $x$, $y$ and $\left|q/p\right|$. It is assumed that there is no direct \CP violation in the CF or DCS decays. It is also assumed that $\Gamma_{12}$, the absorptive part of the off-diagonal element of the mixing matrix, is driven by CF decays. This allows for the interpretation of $\phi$ in terms of $x$, $y$ and $\left|q/p\right|$ using the relations
\begin{equation}
\phi=\arg(y+i\delta x)\;\;{\textrm {and}}\;\;\left|\frac{q}{p}\right|=\sqrt{\frac{1-\delta}{1+\delta}}\;\;\textrm{with}\;\; x=\frac{\Delta m}{\Gamma},\;\;y=\frac{\Delta\Gamma}{2\Gamma},
\end{equation}
leaving only three independent parameters for the fit~\cite{Ciuchini:2007cw,Kagan:2009gb}. The fit can also be performed in terms of $M_{12}$, $\Gamma_{12}$ and $\Phi_{12}=\arg(\Gamma_{12}/M_{12})$ where $M_{12}$ is the dispersive part of the off-diagonal elements of the mixing matrix~\cite{Bevan:2014tha}. The fit average can be summarised as
\begin{equation}
x=(3.5\pm1.5)\times10^{-3},\;\;y=(5.8\pm0.6)\times10^{-3},\;\;\left|\frac{q}{p}\right|-1=(0.7\pm1.8)\times10^{-2},\;\;\phi=(-0.21\pm0.57)^\circ\nonumber
\end{equation}
or alternatively,
\begin{equation}
\left|M_{12}\right|=(4.3\pm1.8)/\textrm{fs},\;\;\left|\Gamma_{12}\right|=(14.1\pm1.4)/\textrm{fs},\;\;\Phi_{12}=(0.8\pm2.6)^\circ\nonumber
\end{equation}
Both $|q/p|$ is compatible with 1 and $\phi$ is compatible with 0 implying the lack of evidence of indirect \CP violation. Moreover, it can be assumed that the SM contribution to $M_{12}$ and $\Gamma_{12}$ is real. However, NP can bring about a phase in $M_{12}$ leading to a non-zero $\Phi_{12}$ while leaving negligible traces in $\Gamma_{12}$. The latter is compatible with 0 from the fit validating the SM hypothesis. All the input used to perform the fit can be found in Table 1 of~\cite{Bevan:2014tha}. The results cited above are from the summer of 2015 and the UTfit collaboration plans on updating the fits soon. The probe of new physics (NP) scales through \DzDzb oscillations assuming $O(1)$ couplings and generic flavour structure were also discussed emphasizing on the fact that these lower bounds on NP scales are second only to those from $\epsilon_K$. The lower bounds can be as high as $O(10^5)$ TeV.

%%%%%%%%%%%%%%
\section{Developments at LHCb}
\label{sec:LHCb}
%\input{lhcb.tex}
%%%%%%%%%%%%%%

Charm particles are among the most abundantly produced in high-energy $pp$ collisions at the Large Hadron Collider. With a $c\bar{c}$ production cross-section (within acceptance) of about 1.5 (3)\,mb at a center-of-mass energy of 7 (13) TeV~\cite{LHCb-PAPER-2012-041,LHCb-PAPER-2015-041}, the LHCb experiment~\cite{Alves:2008zz} is currently not only the ideal place where charm dynamics can be studied, but also the main experimental player.  During the workshop, the LHCb collaboration reported measurements of mixing and \CP violating observables based on an unprecedented large amount of charm-hadron decays that was collected during 2011-2012 (Run\,1) as well as some prospects based on the data that are being collected since 2015 (Run\,2).

A.~Carbone presented the final Run\,1 measurements of direct \CP violation in $D^0\to K^+K^-$ and $D^0\to\pi^+\pi^-$ decays. The analysis measures the time-integrated asymmetries, $A_{\CP}$, in the two final states and their difference $\Delta A_{\CP}\equiv A_{\CP}(\Dz\to K^+K^-)-A_{\CP}(\Dz\to\pi^+\pi^-)$. Results based on samples of \Dz mesons originating from the strong $D^{*+}\to\Dz\pi^+$ decay (pion-tagged decays)~\cite{LHCb-PAPER-2015-055,LHCb-PAPER-2016-035},
\begin{align*}
A_{\CP}(D^0\to K^+K^-) & = \left[+0.14\pm0.15\stat\pm0.10\syst\right]\times10^{-2},\\
A_{\CP}(D^0\to \pi^+\pi^-) & = \left[+0.24\pm0.15\stat\pm0.11\syst\right]\times10^{-2},\\
\Delta A_{\CP} & = \left[-0.10\pm0.08\stat\pm0.03\syst\right]\times10^{-2},
\end{align*}
are combined with those obtained using \Dz mesons produced in inclusive semi-muonic $b$-hadron decays, $\Bbar\to\Dz\mu^-X$, (muon-tagged)~\cite{LHCb-PAPER-2013-003} to give the world's most precise measurements of \CP asymmetries in the charm sector~\cite{LHCb-PAPER-2016-035},
\begin{align*}
A_{\CP}(D^0\to K^+K^-) & = \left[+0.04\pm0.12\stat\pm0.10\syst\right]\times10^{-2},\\
A_{\CP}(D^0\to \pi^+\pi^-) & = \left[+0.07\pm0.14\stat\pm0.11\syst\right]\times10^{-2}.
\end{align*}
No hints of \CP violation are observed and asymmetries at the $10^{-2}$ level seem to be excluded. It is therefore important to broaden the search to experimentally more challenging decay modes where larger effects are still not ruled out. LHCb is moving in this direction by studying channels with neutral final-state particles. A measurement of \CP asymmetries in $D^+_{(s)}\to\eta'\pi^+$ decays, with the $\eta'$ reconstructed in the $\pi^+\pi^-\gamma$ final state, was presented by M. Gersabeck. The \CP asymmetries are measured relative to that of the Cabibbo favoured $D^+\to K_S^0\pi^+$ and $D^+_s\to\phi\pi^+$ control channels. With a total of about $1.1\times10^6$ ($6.6\times10^6$) signal candidates selected for the $D^+_{(s)}$ mode, the obtained results are~\cite{LHCb-PAPER-2016-041}
\begin{align*}
A_{\CP}(D^+\to \eta'\pi^+) & = \left[-0.61\pm0.72\stat\pm0.55\syst\pm0.12(\mathrm{ext.})\right]\times10^{-2},\\
A_{\CP}(D^+_s\to \eta'\pi^+) & = \left[-0.82\pm0.36\stat\pm0.24\syst\pm0.27(\mathrm{ext.})\right]\times10^{-2},
\end{align*}
where the final uncertainties are from the limited precision of the \CP asymmetries of the control channels, which are taken from external inputs~\cite{Ko:2012pe,Abazov:2013woa}. These values are consistent with \CP conservation and improve significantly over the previously best existing measurements performed with data from $e^+e^-$ collisions~\cite{Won:2011ku,Onyisi:2013bjt}. The results show that LHCb can significantly contribute to investigating decays with neutral particles in the final state too.

M.~Gersabeck also discussed experimental techniques to search for direct \CP violation in multi-body charm decays at LHCb, including binned and unbinned methods that can be used to look for differences in the phases-space distribution of charm and anti-charm particles. The highlight of the talk was a measurement of the $D^0\to\pi^+\pi^-\pi^+\pi^-$ decay based on the unbinned, model-independent method called  ``energy test''~\cite{LHCb-PAPER-2016-044}. The method determines and compares the average phase-space distance between candidates of two separate samples (\eg \Dz and \Dzb), in analogy with the measurement of the electrical potential in a volume of mixed positive and negative charges. Two measurements were made: one in which the two samples to be compared were defined purely by the initial neutral $D$ flavour (sensitive to P-even asymmetries), and a second in which the samples were defined by both the $D$ flavour and the sign of a triple-product computed from the pion momenta (sensitive to P-odd asymmetries). The data are found to be consistent with the hypothesis of \CP symmetry with a $p$-values of $(4.6\pm0.5)\times10^{-2}$ and $(0.6\pm0.2)\times10^{-2}$ for the P-even and P-odd tests, respectively. While nothing significant is observed, the P-odd test hints at a possible \CP-violating effect that needs to be investigated with more statistics.

Measurements of mixing and indirect \CP violation in two-body and multi-body charm decays was discussed by K.~Maguire and M.~Martinelli respectively.  Building upon the previous measurements of charm mixing and time-dependent \CP-violation parameters in $D^0\to K^+\pi^+$ decays~\cite{LHCb-PAPER-2013-053}, LHCb performed an updated measurement of the decay-time-dependent ratio of ``wrong-sign'' $D^{*+}\to\Dz(\to K^+\pi^-)\pi^+$ to ``right-sign'' $D^{*+}\to\Dz(\to K^-\pi^+)\pi^+$ rates,
\begin{equation}
R(t) \approx R_D+\sqrt{R_D}\ y'\ \frac{t}{\tau}+\frac{x'^2+y'^2}{4}\left(\frac{t}{\tau}\right)^2,
\end{equation}
where $\tau$ is the average \Dz lifetime. The parameters $x'$ and $y'$ depend linearly on the mixing parameters as $x' \equiv x\cos\delta_{K\pi}+y\sin\delta_{K\pi}$ and $y' \equiv y\cos\delta_{K\pi}-x\sin\delta_{K\pi}$. The parameter $R_D$ and the strong phase $\delta_{K\pi}$ are related to the decay amplitudes as $\mathcal{A}(\Dz\to K^+\pi^-)/\mathcal{A}(\Dzb\to K^+\pi^-) = -\sqrt{R_D} e^{-i\delta_{K\pi}}$. Allowing for \CP violation, the rates $R^+(t)$ and $R^-(t)$ of initially produced \Dz and \Dzb mesons are functions of independent sets of mixing parameters $(R_D^\pm,\, x'^{2\pm},\, y'^\pm)$, where $x'^\pm = |q/p|^{\pm 1}(x'\cos\phi \pm y'\sin\phi)$ and $y'^\pm = |q/p|^{\pm 1}(y'\cos\phi \mp x'\sin\phi)$ are sensitive to indirect \CP violation. The new measurement~\cite{LHCb-PAPER-2016-033} combines the promptly-produced $D^{*+}$ mesons of the previous result with $D^{*+}$ mesons originating from the semi-muonic $\Bbar\to D^{*+}\mu^-X$ decay. While adding only about 2.5\% more signal yield to the pion-tagged sample, these ``doubly-tagged'' candidates feature an improved signal purity and a complementary higher acceptance at low $D$ decay times. As a consequence, the precision on the mixing parameters improved by up to 20\%~\cite{LHCb-PAPER-2016-033}:
\begin{align*}
R_D &= \left(3.553\pm0.054\right)\times10^{-3},\\
y' &= \left(5.23\pm0.84\right)\times10^{-3},\\
x'^2 &= \left(0.36\pm0.43\right)\times10^{-4},
\end{align*}
where the uncertainties include both statistical and systematic contributions. The \Dz and \Dzb mixing rates were found to be consistent with \CP symmetry~\cite{LHCb-PAPER-2016-033}. Indirect \CP violation was also searched for by measuring the asymmetry between the effective decay widths of \Dz and \Dzb mesons decaying to \CP eigenstates, $A_\Gamma$, which is related to the mixing and \CP-violation parameters by
\begin{equation}
A_\Gamma \approx \frac{y}{2}\left(\left|\frac{q}{p}\right|-\left|\frac{p}{q}\right|\right)\cos\phi-\frac{x}{2}\left(\left|\frac{q}{p}\right|+\left|\frac{p}{q}\right|\right)\sin\phi.
\end{equation}
The measurement is based on pion-tagged decays and is performed with two independent methods yielding consistent results~\cite{LHCb-PAPER-2016-063}:
\begin{align*}
A_\Gamma(D^0\to K^+K^-) & = \left[-0.30\pm0.22\stat\pm0.10\syst\right]\times10^{-3},\\
A_\Gamma(D^0\to \pi^+\pi^-) & = \left[+0.46\pm0.58\stat\pm0.12\syst\right]\times10^{-3}.
\end{align*}
The results show no evidence for \CP violation and improve on the precision of the previous best determinations by nearly a factor of two.

In addition to measurements of mixing and indirect \CP violation using two-body decays, LHCb has also presented the first observation of charm mixing using $D^0\to K^+\pi^-\pi^+\pi^-$ decays~\cite{LHCb-PAPER-2015-057}. While the current measurement is still not competitive in constraining the mixing parameters with the results based on two-body decays, this multi-body mode, together with the $D^0\to K_S^0\pi^+\pi^+$ channel, is expected to play an important role in future analyses using Run\,2 data. M.~Martinelli showed, indeed, that due to the increased production cross-section and improved detector performances, LHCb is now collecting data on these decays at a rate that is 4 to 6 times larger than what was achieved during Run\,1.

An overview of mixing and \CP violation measurement in two body decays of charm at the LHCb can be found in~\cite{Maguire:2017hsw}.
%%%%%%%%%%%%%%
\section{Status and Prospects at BESIII}
\label{sec:BESIII}
%\input{BESIII.tex}
%%%%%%%%%%%%%%

The BESIII experiment has been collecting data from $e^+e^-$ collisions of the BEPC accelerator since 2009.
Data are taken at different energies, including the $\Psi(3770)$ that
decays in a quantum correlated \DzDzb  state which allows for access to the strong phases between the $D$ and $\Dbar$
decays. BESIII measurements of the strong phase
difference between $D$ and $\Dbar$ decays (in bins of the Dalitz plot for multi-body decays) and the coherent factors are fundamental
to interpreting Belle\,II and LHCb charm and $B$ measurements in terms of SM parameters. From the point of view of statistics, the $B$-Factories and LHCb have much larger data samples, therefore BESIII 
is not competitive on measurements accessible to the former two.
X.~R.~Lyu presented several measurements, many of which will play an important
role for the interpretation of LHCb~\cite{Malde:2223391} and Belle\,II results.

X.~R.~Lyu reported on the measurement of the strong phase between the \Dz and the
\Dzb decaying in the $K\pi$ final state, $\cos\delta_{K\pi} = 1.02 \pm 0.11 \pm0.06 \pm 0.01$ measured with
2.93 fb$^{-1}$ of data at 3.773 GeV center of mass energy~\cite{Ablikim2014227}. This measurement is the most precise till date. The expected precision with 10 fb$^{-1}$ of data will reach 0.07.
When the \Dz decays to final states with more than two particles the
strong phase depends on the Dalitz plot variables. For the golden channel
for measurement of the mixing and the \CP violation
parameters is $\Dz \to K^0_S\pi^+\pi^-$, X.~R.~Lyu
presented a preliminary study of the measurement of the sine ($s_i$) and
cosine ($c_i$) of the strong-phase difference averaged in each Dalitz plot bin ($i$) and weighted by the absolute decay rate, with 2.93
fb$^{-1}$ of data at 3.773 GeV.
The measurement of $c_i$ and $s_i$ is fundamental in the
model-independent measurement of $\gamma/\phi_3$ in the channel
$B^{\pm}\to DK^{\pm}$ with $D\to K^0_S\pi^+\pi^-$.
Only the statistical error was included, and a reduction of $40\%$ on the error of the CKM angle is predicted.
Other channels are under study ($\Dz\to K^0_S K^+
K^-$, $\Dz\to K^{\pm} \pi^{\mp} \pi^+\pi^-$, $\Dz\to K^{\pm} \pi^{\mp}
\pi^0$,  $\Dz\to K^{\pm} K^{\mp} \pi^0$) and some others are planned. All
of them are of great interest for Belle\,II and LHCb.

Preliminary results of time-integrated \CP asymmetry in the SCS mode $D^+ \to K_{S,L} K^+(\pi^0)$
decays were presented. The measurements are
all dominated by statistical error, ranging between 3\% and 4\%,
with a systematic error between 1\% and 2\%. Preliminary results on
branching fractions measurements have also been shown for many channels.
The measurement of the branching fractions is an important ingredient for the
validation of the $SU(3)_{FL}$-based models discussed in section~\ref{sec:pheno:twobody}.

Finally, X.~R.~Lyu presented the current and expected uncertainties on
$|V_{cs}|$, $|V_{cd}|$ and on the $D$ form factors. All of them are expected to reach the sub-\% level precision in the future, with improvements on the LQCD calculations (for the CKM parameters) and increase of
statistics collected by BESIII to 10 fb$^{-1}$.

%%%%%%%%%%%%%%
\section{Recent Results at the $B$-Factories and Prospects at Belle\,II}
\label{sec:BFactories}
%\input{bfactories.tex}
%%%%%%%%%%%%%%

The $B$-Factories have always played a central role in charm
physics. Belle and \babar\ have together accumulated  roughly 1.5~ab$^{-1}$ of data. Belle\,II will accumulate 50~ab$^{-1}$ of data, significantly increasing
 statistics and reducing the errors on measurements.
With respect to the hadron machines, the B-Factories offer a clean
environment with a very high trigger efficiency, excellent neutral particle
identification, high flavour-tagging efficiency with low dilution and the
possibility to do missing energy analyses. They pay the price with a smaller
cross section and consequently smaller data samples. Given the fact
that the systematic
errors are quite different from those at LHCb and many measurements are
overlapping, $B$-Factories
and LHCb can be considered as complementary experiments, and both of them are crucial
to search for beyond SM physics in the charm sector.

Charm measurements at Belle\,II will benefit not only from the increase of
statistics, but also from the improved reconstruction performances of
the detector and the software. A.~Schwartz showed that the \Dz proper
time resolution is a factor two better at Belle\,II than \babar, due
to the improved vertexing detector and tracking performances. He also showed
that this resolution is achieved both for {\Dz}s coming from the $D^*$
decay and for prompt {\Dz}s. S. Bahinipati reported on a new
flavour-tagging technique that will allow to tag prompt {\Dz}s by
reconstructing the charged kaons in the rest-frame of the event. This new
technique has been fully characterized by simulated events, and it
promises an equivalent increase of luminosity of around 35\%.

V. Bhardwaj reported on recent results obtained by Belle and
\babar. He reported on the measurement of $x'^2$, $y'$ and $R_D$
by Belle with the complete dataset of 976~fb$^{-1}$~\cite{PhysRevLett.112.111801}, with errors roughly
twice as large as the ones obtained by LHCb. A. Schwartz presented
a Monte Carlo study on this channel that allows for the estimate of the
expected precision on the mixing parameters with the improved proper
time resolution at Belle\,II. The preliminary results allowing for \CP violation are
reported in Table~\ref{tab:toy}, and represent a significant
improvement with respect to the Belle value scaled with luminosity.
\begin{table}
\begin{center}
\begin{tabular}{l|c|c|c}
observable & 5 ab$^{-1}$& 20 ab$^{-1}$ & 50 ab$^{-1}$\\
\hline
$x'$ (\%) & 0.37 & 0.23 & 0.15\\
$y'$ (\%)& 0.26 & 0.17 & 0.10\\
$|q/p|$ & 0.197& 0.089& 0.051\\
$\phi$ (deg)& 15.5 & 9.2 &5.7\\
\end{tabular}
\caption[]{Expected error on mixing and \CP violation parameters obtained
by a ToyMC study including the improved proper time error resolution.}
\label{tab:toy}
\end{center}
\end{table}

V. Bhardwaj discussed the Belle time-dependent
analysis using the Dalitz plot of the decays $D^0\to K_S \pi^+ \pi^-$,
directly sensitive to the mixing parameters, $|q/p|$ and its phase~\cite{PhysRevD.89.091103}.
The Dalitz plot distribution is described by a model that includes
several resonances, whose amplitudes and phases are extracted from data.
The analysis is the most precise single measurement of the mixing
parameters, yielding 
$$x~=~0.56~\pm~0.19~^{+0.03+0.06}_{-0.09 - 0.09}, \quad
y~=~0.30~\pm~0.15~^{+0.04+0.03}_{-0.05 - 0.06},$$
$$\left|\frac{q}{p}\right|~=~0.90~^{+0.16+0.05+0.06}_{-0.15-0.04 - 0.05}, \quad
\arg\left(\frac{q}{p}\right)~=~6~\pm~11~^{+3+3}_{-4 -4},$$
where the last error is the one related to the Dalitz plot model.
A. Schwartz showed that the measurement at Belle II will be limited
by the Dalitz plot model related error, that
does not completely scale with luminosity. It will therefore be
important to use a model-independent approach, using the sines and
cosines of the strong phases measured by BESIII.
He also presented a Toy MonteCarlo study similar to the one
discussed above for a Dalitz Analysis of the channel $D^0\to K^+ \pi^-
\pi^0$. Considering only the statistical error, the expected
precision on the rotated mixing parameters $x''$ and $y''$ with 50~ab$^{-1}$ of data are 0.057\% and 0.049\% respectively.

V. Bhardwaj discussed recent time-integrated \CP asymmetries
measured at Belle. As mentioned above, an interesting channel for the discovery of \CP violation in the charm sector is
$D^0\to K^0_S K^0_S$, since the estimated \CP asymmetry may reach 1\%.
Belle preliminary result was presented with 
$$A_{\CP}(D^0\to K^0_S K^0_S) = (-0.02\pm 1.53 \pm 0.17)\%.$$ 
S. Bahinipati showed that both the
statistical and the systematic error on this measurement nicely scale
with luminosity, with a potential precision of 0.2\% with 50~ab$^{-1}$ of data,
which could be enough to find the first evidence of \CP violation in charm.
V. Bhardwaj  and S. Bahinipati discussed the Belle measurement of
\CP asymmetry at Belle~\cite{PhysRevLett.112.211601} and the prospects at Belle\,II for the channels
$D^0 \to \pi^0 \pi^0$: 
$$A_{\CP}(D^0\to\pi^0\pi^0) = (-0.03\pm0.64\pm0.10)\%$$
with an expected precision with 50~ab$^{-1}$ of data of 0.09\%, and for  $D^0 \to K^0_S \pi^0$: 
$$A_{\CP}(D^0 \to K^0_S \pi^0) = (-0.21\pm0.16\pm0.07)\%$$
with an expected precision with 50~ab$^{-1}$ of data of 0.03\%.

Finally, a recent Belle measurement~\cite{PhysRevLett.118.051801} of radiative decay of the $D^0$ into
a vector was discussed. The first observation of $D^0\to \rho^0
\gamma$ was reported and the measurement of the \CP asymmetries in 
$D^0\to \phi^0 \gamma$ , $D^0\to \Kbar^{*0} \gamma$ and $D^0\to \rho^0 \gamma$ 
were presented, with statistical error of 0.066, 0.020 and 0.151
respectively and negligible systematic errors. S. Bahinipati showed
the predicted precision expected at Belle\,II, arguing that the error
will primarily scale with luminosity, obtaining:  0.02, 0.01 and
0.003 respectively.

An overview of the results at Belle and the prospects of measurements at Belle II can be found in~\cite{Bahinipati:2017hsw,Bhardwaj:2017hsw}

%%%%%%%%%%%%%%
\section{Conclusion}
\label{sec:conclusion}
%%%%%%%%%%%%%%
The sessions of the WG7 saw a comprehensive review of \CP violations and oscillations in the charm mesons system covering both the phenomenological and experimental aspects.

From the phenomenological side, it has been underlined that the greatest hurdle in predicting \CP asymmetries in $D\to P P$ decays lies in the estimation of the penguin amplitudes which are non-local and hence not under theoretical control. This can be side-stepped by appealing to sum rules between \CP asymmetries. One particular channel of interest for \CP asymmetry is $D^0\to K^0_SK^0_S$ where the suppression of the \CP conserving part due to an approximate $SU(3)_{FL}$ symmetry enhances the asymmetry. Beyond, two-body decay channels, much can be gleaned from three- and four-body decay modes. Amplitude analysis holds the key to the measurement of strong phases and local \CP asymmetries although they require large statistics. Besides direct \CP asymmetries, measurement indirect \CP violation in the neutral $D$ system has seen some progress. The averaging performed by the UTfit Collaboration shows that results are compatible with the no indirect \CP violation hypothesis.

During Run\,1, LHCb has collected what is by far the largest sample of charm hadron decays currently available. Most of the presented results are based on this sample and in many cases have reached sensitivities close to the naive SM expectations. Direct and indirect \CP asymmetries are now measured in $D^0\to K^+K^-$ and $D^0\to\pi^+\pi^-$ to be consistent with zero with uncertainties well below 1\%. LHCb has also demonstrated to be competitive with decay to final state involving neutral particles, such as $D^+_{(s)}\to \eta'(\to\pi^+\pi^-\gamma)\pi^+$, where \CP asymmetries are measured at the sub-\% level. While Run\,1 data is still being fully exploited, substantial improvements are expected from the analysis of Run\,2 data, particularly for multi-body final states.

Running at the charm threshold, BESIII has produced quite a few interesting results. The importance of many of these lie in the fact that they will provide important input for many searches in LHCb~\cite{Malde:2223391} and Belle\,II. The latter will also benefit from a better detector and improved reconstruction performances in addition to a significantly larger statistical sample. With a recent measurement of \CP asymmetry in $D^0\to K^0_SK^0_S$ with the Belle data, Belle\,II seems well geared for this measurement with precision comparable with the current SM prediction.

%%%%%%%%%%%%%%
\section{Acknowledgement}
\label{sec:ack}
%%%%%%%%%%%%%%
We would like to thank the organizers of CKM 2016 for giving us the
opportunity of working on the committee for WG7. A. P. would like to
acknowledge the support from ERC Ideas Starting Grant n.~279972
``NPFlavour''. G. C. would like to acknowledge the support from the
Alexander von Humboldt Foundation.

\bibliographystyle{JHEP}
\bibliography{CKM2016_WG7}

\providecommand{\href}[2]{#2}\begingroup\raggedright\begin{thebibliography}{10}

\bibitem{Malde:2223391}
{\scshape LHCb} collaboration, S.~S. Malde et~al., ``{Synergy of BESIII and
  LHCb physics programmes}.'' Oct, 2016.

\bibitem{Hiller:2012xm}
G.~Hiller, M.~Jung and S.~Schacht, \emph{{SU(3)-flavor anatomy of nonleptonic
  charm decays}},
  \href{http://dx.doi.org/10.1103/PhysRevD.87.014024}{\emph{Phys. Rev.}
  {\bfseries D87} (2013) 014024} DO-TH-12-22,
  [\href{https://arxiv.org/abs/1211.3734}{{\ttfamily 1211.3734}}].

\bibitem{Muller:2015rna}
S.~M{\"u}ller, U.~Nierste and S.~Schacht, \emph{{Sum Rules of Charm CP
  Asymmetries beyond the SU(3)$_F$ Limit}},
  \href{http://dx.doi.org/10.1103/PhysRevLett.115.251802}{\emph{Phys. Rev.
  Lett.} {\bfseries 115} (2015) 251802} TTP15-020,
  [\href{https://arxiv.org/abs/1506.04121}{{\ttfamily 1506.04121}}].

\bibitem{Nierste:2015zra}
U.~Nierste and S.~Schacht, \emph{{CP Violation in $D^0\rightarrow K_SK_S$}},
  \href{http://dx.doi.org/10.1103/PhysRevD.92.054036}{\emph{Phys. Rev.}
  {\bfseries D92} (2015) 054036} TTP15-027,
  [\href{https://arxiv.org/abs/1508.00074}{{\ttfamily 1508.00074}}].

\bibitem{Muller:2015lua}
S.~M{\"u}ller, U.~Nierste and S.~Schacht, \emph{{Topological amplitudes in $D$
  decays to two pseudoscalars: A global analysis with linear $SU(3)_F$
  breaking}}, \href{http://dx.doi.org/10.1103/PhysRevD.92.014004}{\emph{Phys.
  Rev.} {\bfseries D92} (2015) 014004} TTP15-015,
  [\href{https://arxiv.org/abs/1503.06759}{{\ttfamily 1503.06759}}].

\bibitem{Aaij:2015fua}
{\scshape LHCb} collaboration, R.~Aaij et~al., \emph{{Measurement of the
  time-integrated $CP$ asymmetry in $D^0 \to K^0_S K^0_S$ decays}},
  \href{http://dx.doi.org/10.1007/JHEP10(2015)055}{\emph{JHEP} {\bfseries 10}
  (2015) 055} LHCB-PAPER-2015-030, CERN-PH-EP-2015-215,
  LHCB-PAPER-2015-030-CERN-PH-EP-2015-215,
  [\href{https://arxiv.org/abs/1508.06087}{{\ttfamily 1508.06087}}].

\bibitem{Abdesselam:2016gqq}
A.~Abdesselam et~al., \emph{{Measurement of $CP$ asymmetry in the $D^{0} \to
  K^0_S K^0_S$ decay at Belle}},
  \href{https://arxiv.org/abs/1609.06393}{{\ttfamily 1609.06393}}
  BELLE-CONF-1609, [\href{https://arxiv.org/abs/1609.06393}{{\ttfamily
  1609.06393}}].

\bibitem{Nierste:2017hsw}
U.~Nierste and S.~Schacht, \emph{{CP asymmetries in D decays to two
  pseudoscalars}}, {\emph{PoS} {\bfseries CKM2016} (2017) 134}.

\bibitem{Amato:2016xjv}
J.~H. Alvarenga~Nogueira et~al., \emph{{Summary of the 2015 LHCb workshop on
  multi-body decays of D and B mesons}},
  \href{https://arxiv.org/abs/1605.03889}{{\ttfamily 1605.03889}}.

\bibitem{LHCb-CONF-2016-008}
{\scshape LHCb Collaboration} collaboration, \emph{{Dalitz plot analysis of the
  $D^+ \rightarrow K^- K^+ K^+$ decay with the isobar model }},
  LHCb-CONF-2016-008.

\bibitem{Aaij:2015lsa}
{\scshape LHCb} collaboration, R.~Aaij et~al., \emph{{Studies of the resonance
  structure in $D^0\to K^0_S K^{\pm}\pi^{\mp}$ decays}},
  \href{http://dx.doi.org/10.1103/PhysRevD.93.052018}{\emph{Phys. Rev.}
  {\bfseries D93} (2016) 052018} CERN-PH-EP-2015-238, LHCB-PAPER-2015-026,
  [\href{https://arxiv.org/abs/1509.06628}{{\ttfamily 1509.06628}}].

\bibitem{Peng:2014oda}
{\scshape Belle} collaboration, T.~Peng et~al., \emph{{Measurement of
  $D^0-\bar{D}^0$ mixing and search for indirect CP violation using $D^0\to
  K_S^0\pi^+\pi^-$ decays}},
  \href{http://dx.doi.org/10.1103/PhysRevD.89.091103}{\emph{Phys. Rev.}
  {\bfseries D89} (2014) 091103},
  [\href{https://arxiv.org/abs/1404.2412}{{\ttfamily 1404.2412}}].

\bibitem{Aaltonen:2012nd}
{\scshape CDF} collaboration, T.~Aaltonen et~al., \emph{{Measurement of
  CP-violation asymmetries in $D^0 \to K_S \pi^+ \pi^-$}},
  \href{http://dx.doi.org/10.1103/PhysRevD.86.032007}{\emph{Phys. Rev.}
  {\bfseries D86} (2012) 032007} FERMILAB-PUB-12-339-E,
  [\href{https://arxiv.org/abs/1207.0825}{{\ttfamily 1207.0825}}].

\bibitem{Lees:2012nn}
{\scshape BaBar} collaboration, J.~P. Lees et~al., \emph{{Search for direct CP
  violation in singly Cabibbo-suppressed $D^\pm \to K^+K^-\pi^\pm$ decays}},
  \href{http://dx.doi.org/10.1103/PhysRevD.87.052010}{\emph{Phys. Rev.}
  {\bfseries D87} (2013) 052010} BABAR-PUB-12-014, SLAC-PUB-15077,
  [\href{https://arxiv.org/abs/1212.1856}{{\ttfamily 1212.1856}}].

\bibitem{Aaij:2015xoa}
{\scshape LHCb} collaboration, R.~Aaij et~al., \emph{{Model-independent
  measurement of mixing parameters in $D^{0} \to K_{S}^{0}\pi^{+}\pi^{-}$
  decays}}, \href{http://dx.doi.org/10.1007/JHEP04(2016)033}{\emph{JHEP}
  {\bfseries 04} (2016) 033} CERN-PH-EP-2015-249, LHCB-PAPER-2015-042,
  [\href{https://arxiv.org/abs/1510.01664}{{\ttfamily 1510.01664}}].

\bibitem{dArgent:2016rbp}
P.~d'Argent, J.~Benton, J.~Dalseno, E.~Gersabeck, S.~Harnew, P.~Naik et~al.,
  \emph{{Amplitude analysis of $D^{0} \rightarrow \pi^{+} \pi^{-} \pi^{+}
  \pi^{-}$ decays using CLEO-c data}}, {\emph{PoS} {\bfseries CHARM2016} (2016)
  084}, [\href{https://arxiv.org/abs/1611.09253}{{\ttfamily 1611.09253}}].

\bibitem{Aaij:2014qwa}
{\scshape LHCb} collaboration, R.~Aaij et~al., \emph{{Search for $CP$ violation
  using $T$-odd correlations in $D^0 \to K^+K^-\pi^+\pi^-$ decays}},
  \href{http://dx.doi.org/10.1007/JHEP10(2014)005}{\emph{JHEP} {\bfseries 10}
  (2014) 005} CERN-PH-EP-2014-194, LHCB-PAPER-2014-046,
  [\href{https://arxiv.org/abs/1408.1299}{{\ttfamily 1408.1299}}].

\bibitem{delAmoSanchez:2010xj}
{\scshape BaBar} collaboration, P.~del Amo~Sanchez et~al., \emph{{Search for CP
  violation using $T$-odd correlations in $D^0 \to K^+ K^- \pi^+ \pi^-$
  decays}}, \href{http://dx.doi.org/10.1103/PhysRevD.81.111103}{\emph{Phys.
  Rev.} {\bfseries D81} (2010) 111103} BABAR-PUB-09-039, SLAC-PUB-13996,
  [\href{https://arxiv.org/abs/1003.3397}{{\ttfamily 1003.3397}}].

\bibitem{Ciuchini:2007cw}
M.~Ciuchini, E.~Franco, D.~Guadagnoli, V.~Lubicz, M.~Pierini, V.~Porretti
  et~al., \emph{{$D - \bar{D}$ mixing and new physics: General considerations
  and constraints on the MSSM}},
  \href{http://dx.doi.org/10.1016/j.physletb.2007.08.055}{\emph{Phys. Lett.}
  {\bfseries B655} (2007) 162--166},
  [\href{https://arxiv.org/abs/hep-ph/0703204}{{\ttfamily hep-ph/0703204}}].

\bibitem{Kagan:2009gb}
A.~L. Kagan and M.~D. Sokoloff, \emph{{On Indirect CP Violation and
  Implications for $D^0 - \bar{D}^0$ and $B_s - \bar{B}_s$ mixing}},
  \href{http://dx.doi.org/10.1103/PhysRevD.80.076008}{\emph{Phys. Rev.}
  {\bfseries D80} (2009) 076008},
  [\href{https://arxiv.org/abs/0907.3917}{{\ttfamily 0907.3917}}].

\bibitem{Bevan:2014tha}
{\scshape UTfit} collaboration, A.~J. Bevan et~al., \emph{{The UTfit
  collaboration average of $D$ meson mixing data: Winter 2014}},
  \href{http://dx.doi.org/10.1007/JHEP03(2014)123}{\emph{JHEP} {\bfseries 03}
  (2014) 123}, [\href{https://arxiv.org/abs/1402.1664}{{\ttfamily 1402.1664}}].

\bibitem{LHCb-PAPER-2012-041}
{\scshape LHCb} collaboration, R.~Aaij et~al., \emph{{Prompt charm production
  in $pp$ collisions at $\sqrt{s}=7$\,TeV}},
  \href{http://dx.doi.org/10.1016/j.nuclphysb.2013.02.010}{\emph{Nucl. Phys.}
  {\bfseries B871} (2013) 1} LHCb-PAPER-2012-041, CERN-PH-EP-2013-009,
  [\href{https://arxiv.org/abs/1302.2864}{{\ttfamily 1302.2864}}].

\bibitem{LHCb-PAPER-2015-041}
{\scshape LHCb} collaboration, R.~Aaij et~al., \emph{{Measurements of prompt
  charm production cross-sections in $pp$ collisions at $\sqrt{s} = 13$\,TeV}},
  \href{http://dx.doi.org/10.1007/JHEP03(2016)159}{\emph{JHEP} {\bfseries 03}
  (2016) 159} {LHCb-PAPER-2015-041, CERN-PH-EP-2015-272},
  [\href{https://arxiv.org/abs/1510.01707}{{\ttfamily 1510.01707}}].

\bibitem{Alves:2008zz}
{\scshape LHCb} collaboration, A.~A. Alves~Jr. et~al., \emph{{The LHCb detector
  at the LHC}},
  \href{http://dx.doi.org/10.1088/1748-0221/3/08/S08005}{\emph{JINST}
  {\bfseries 3} (2008) S08005}.

\bibitem{LHCb-PAPER-2015-055}
{\scshape LHCb} collaboration, R.~Aaij et~al., \emph{{Measurement of the
  difference of time-integrated \CP asymmetries in $\Dz\to K^-K^+$ and
  $\Dz\to\pi^-\pi^+$ decays}},
  \href{http://dx.doi.org/10.1103/PhysRevLett.116.191601}{\emph{Phys. Rev.
  Lett.} {\bfseries 116} (2016) 191601} {LHCb-PAPER-2015-055,
  CERN-EP-2016-022}, [\href{https://arxiv.org/abs/1602.03160}{{\ttfamily
  1602.03160}}].

\bibitem{LHCb-PAPER-2016-035}
{\scshape LHCb} collaboration, R.~Aaij et~al., \emph{{Measurement of \CP
  asymmetry in $\Dz \to K^+K^-$ decays}},
  \href{https://arxiv.org/abs/1610.09467}{{\ttfamily 1610.09467}}
  {LHCb-PAPER-2016-035, CERN-EP-2016-259},
  [\href{https://arxiv.org/abs/1610.09467}{{\ttfamily 1610.09467}}].

\bibitem{LHCb-PAPER-2013-003}
{\scshape LHCb} collaboration, R.~Aaij et~al., \emph{{Search for direct \CP
  violation in $\Dz\to h^- h^+$ modes using semileptonic $B$ decays}},
  \href{http://dx.doi.org/10.1016/j.physletb.2013.04.061}{\emph{Phys. Lett.}
  {\bfseries B723} (2013) 33} CERN-PH-EP-2013-039, LHCb-PAPER-2013-003,
  [\href{https://arxiv.org/abs/1303.2614}{{\ttfamily 1303.2614}}].

\bibitem{LHCb-PAPER-2016-041}
{\scshape LHCb} collaboration, R.~Aaij et~al., \emph{{Measurement of \CP\
  asymmetries in $D^\pm \to \eta'\pi^\pm$ and $D_s^\pm \to\eta'\pi^\pm$
  decays}},  \href{https://arxiv.org/abs/1701.01871}{{\ttfamily 1701.01871}}
  LHCB-PAPER-2016-041, CERN-EP-2016-315,
  [\href{https://arxiv.org/abs/1701.01871}{{\ttfamily 1701.01871}}].

\bibitem{Ko:2012pe}
{\scshape Belle} collaboration, B.~R. Ko et~al., \emph{{Evidence for CP
  Violation in the Decay $D^+\to K^0_S\pi^+$}},
  \href{http://dx.doi.org/10.1103/PhysRevLett.109.021601}{\emph{Phys. Rev.
  Lett.} {\bfseries 109} (2012) 021601},
  [\href{https://arxiv.org/abs/1203.6409}{{\ttfamily 1203.6409}}].

\bibitem{Abazov:2013woa}
{\scshape D0} collaboration, V.~M. Abazov et~al., \emph{{Measurement of the
  direct CP-violating charge asymmetry in ${D_s^\pm \rightarrow \phi
  \pi^{\pm}}$ decays}},
  \href{http://dx.doi.org/10.1103/PhysRevLett.112.111804}{\emph{Phys. Rev.
  Lett.} {\bfseries 112} (2014) 111804} FERMILAB-PUB-13-550-E,
  [\href{https://arxiv.org/abs/1312.0741}{{\ttfamily 1312.0741}}].

\bibitem{Won:2011ku}
{\scshape Belle} collaboration, E.~Won et~al., \emph{{Observation of $D^+
  \rightarrow K^{+} \eta^{(\prime)}$ and Search for CP Violation in $D^+
  \rightarrow \pi^+ \eta^{(\prime)}$ Decays}},
  \href{http://dx.doi.org/10.1103/PhysRevLett.107.221801}{\emph{Phys. Rev.
  Lett.} {\bfseries 107} (2011) 221801},
  [\href{https://arxiv.org/abs/1107.0553}{{\ttfamily 1107.0553}}].

\bibitem{Onyisi:2013bjt}
{\scshape CLEO} collaboration, P.~U.~E. Onyisi et~al., \emph{{Improved
  Measurement of Absolute Hadronic Branching Fractions of the $D_s^+$ Meson}},
  \href{http://dx.doi.org/10.1103/PhysRevD.88.032009}{\emph{Phys. Rev.}
  {\bfseries D88} (2013) 032009} CLNS-13-2086, CLEO-13-01,
  [\href{https://arxiv.org/abs/1306.5363}{{\ttfamily 1306.5363}}].

\bibitem{LHCb-PAPER-2016-044}
{\scshape LHCb} collaboration, R.~Aaij et~al., \emph{{Search for \CP violation
  in the phase space of $\Dz\to \pi^+\pi^-\pi^+\pi^-$ decays}},
  \href{https://arxiv.org/abs/1612.03207}{{\ttfamily 1612.03207}}
  {LHCb-PAPER-2016-044, CERN-EP-2016-287},
  [\href{https://arxiv.org/abs/1612.03207}{{\ttfamily 1612.03207}}].

\bibitem{LHCb-PAPER-2013-053}
{\scshape LHCb} collaboration, R.~Aaij et~al., \emph{{Measurement of
  $\Dz$--$\Dzb$ mixing parameters and search for \CP violation using $\Dz\to
  K^+\pi^-$ decays}},
  \href{http://dx.doi.org/10.1103/PhysRevLett.111.251801}{\emph{Phys. Rev.
  Lett.} {\bfseries 111} (2013) 251801} CERN-PH-EP-2013-176,
  LHCb-PAPER-2013-053, [\href{https://arxiv.org/abs/1309.6534}{{\ttfamily
  1309.6534}}].

\bibitem{LHCb-PAPER-2016-033}
{\scshape LHCb} collaboration, R.~Aaij et~al., \emph{{Measurement of charm
  mixing and \CP violation using $\Dz\to K^\pm\pi^\mp$ decays}},
  \href{https://arxiv.org/abs/1611.06143}{{\ttfamily 1611.06143}}
  LHCb-PAPER-2016-033, CERN-EP-2016-280,
  [\href{https://arxiv.org/abs/1611.06143}{{\ttfamily 1611.06143}}].

\bibitem{LHCb-PAPER-2016-063}
{\scshape LHCb} collaboration, R.~Aaij et~al., \emph{{Measurement of the \CP
  violation parameter $A_\Gamma$ in $D^0 \to K^+K^-$ and $D^0 \to \pi^+\pi^-$
  decays}},  \href{https://arxiv.org/abs/1702.06490}{{\ttfamily 1702.06490}}
  LHCB-PAPER-2016-063, CERN-EP-2017-028,
  [\href{https://arxiv.org/abs/1702.06490}{{\ttfamily 1702.06490}}].

\bibitem{LHCb-PAPER-2015-057}
{\scshape LHCb} collaboration, R.~Aaij et~al., \emph{{First observation of
  $\Dz$--$\Dzb$ oscillations in $\Dz\to K^+\pi^-\pi^+\pi^-$ decays and a
  measurement of the associated coherence parameters}},
  \href{http://dx.doi.org/10.1103/PhysRevLett.116.241801}{\emph{Phys. Rev.
  Lett.} {\bfseries 116} (2016) 241801} {LHCb-PAPER-2015-057,
  CERN-EP-2016-021}, [\href{https://arxiv.org/abs/1602.07224}{{\ttfamily
  1602.07224}}].

\bibitem{Maguire:2017hsw}
K.~Maguire, \emph{{Measurements of mixing and indirect CP violation in two-body
  charm decays at LHCb}}, {\emph{PoS} {\bfseries CKM2016} (2017) 140}.

\bibitem{Ablikim2014227}
{\scshape BESIII} collaboration, M.~Ablikim et~al., \emph{{Measurement of the
  $D\to K^-\pi^+$ strong phase difference in $\psi(3770)\to
  D^0\overline{D}{}^0$}},
  \href{http://dx.doi.org/10.1016/j.physletb.2014.05.071}{\emph{Phys. Lett.}
  {\bfseries B734} (2014) 227--233},
  [\href{https://arxiv.org/abs/1404.4691}{{\ttfamily 1404.4691}}].

\bibitem{PhysRevLett.112.111801}
{\scshape Belle Collaboration} collaboration, B.~R. Ko et~al.,
  \emph{Observation of {\DzDzb} mixing in $e^+e^-$ collisions},
  \href{http://dx.doi.org/10.1103/PhysRevLett.112.111801}{\emph{Phys. Rev.
  Lett.} {\bfseries 112} (Mar, 2014) 111801}.

\bibitem{PhysRevD.89.091103}
{\scshape Belle Collaboration} collaboration, T.~Peng et~al., \emph{Measurement
  of {\DzDzb} mixing and search for indirect ${CP}$ violation using
  ${D}^{0}\ensuremath{\rightarrow}{K}_{S}^{0}{\ensuremath{\pi}}^{+}{\ensuremath{\pi}}^{\ensuremath{-}}$
  decays}, \href{http://dx.doi.org/10.1103/PhysRevD.89.091103}{\emph{Phys. Rev.
  D} {\bfseries 89} (May, 2014) 091103}.

\bibitem{PhysRevLett.112.211601}
N.~K. Nisar et~al., \emph{Search for ${CP}$ violation in
  ${D}^{0}\ensuremath{\rightarrow}{\ensuremath{\pi}}^{0}{\ensuremath{\pi}}^{0}$
  decays}, \href{http://dx.doi.org/10.1103/PhysRevLett.112.211601}{\emph{Phys.
  Rev. Lett.} {\bfseries 112} (May, 2014) 211601}.

\bibitem{PhysRevLett.118.051801}
{\scshape Belle Collaboration} collaboration, T.~Nanut et~al.,
  \emph{Observation of
  ${D}^{0}\ensuremath{\rightarrow}{\ensuremath{\rho}}^{0}\ensuremath{\gamma}$
  and search for {CP} violation in radiative charm decays},
  \href{http://dx.doi.org/10.1103/PhysRevLett.118.051801}{\emph{Phys. Rev.
  Lett.} {\bfseries 118} (Jan, 2017) 051801}.

\bibitem{Bahinipati:2017hsw}
S.~Bahinipati, \emph{{Prospects on time-integrated CPV measurements at Belle
  II}}, {\emph{PoS} {\bfseries CKM2016} (2017) 135}.

\bibitem{Bhardwaj:2017hsw}
V.~Bhardwaj, \emph{{Latest Charm Mixing and $CP$ results from $B$-factories}},
  {\emph{PoS} {\bfseries CKM2016} (2017) 139},
  [\href{https://arxiv.org/abs/1703.04397}{{\ttfamily 1703.04397}}].

\end{thebibliography}\endgroup

\end{document}